\documentclass[11pt]{amsart}
\usepackage{amssymb}
\usepackage{graphicx}
\usepackage{amscd}

\newtheorem{theorem}{Theorem}
\theoremstyle{plain}

\newtheorem{definition}{Definition}

\newtheorem{lemma}{Lemma}

\newtheorem{problem}{Problem}

\newtheorem{remark}{Remark}

\numberwithin{equation}{section}
\numberwithin{theorem}{section}
\numberwithin{lemma}{section}
\numberwithin{proposition}{section}
\numberwithin{corollary}{section}
\numberwithin{definition}{section}

\input{tcilatex}

\begin{document}
\title[Wave-type pseudo-differential equations ]{Decay of solutions of
wave-type pseudo-differential equations over $p-$adic fields}
\author{W. A. Zuniga-Galindo}
\address{ Department of Mathematics and Computer Science, Barry University,
11300 N.E. Second Avenue, Miami Shores, Florida 33161, USA.}

\begin{abstract}
We show that the solutions of $p-$adic pseudo-differential \ equations of
wave type have a decay similar to the solutions of \ classical generalized
wave equations.
\end{abstract}

\thanks{Project sponsored by the National Security Agency under Grant Number
H98230-06-1-0040. The United States Government is authorized to reproduce
and distribute reprints notwithstanding any copyright notation herein.}
\email{wzuniga@mail.barry.edu}
\subjclass{Primary 35S99, 47S10; Secondary 11S40}
\keywords{Non-archimedean pseudo-differential equations, restriction of
Fourier transforms, exponential sums mdulo $p^{m}$, Igusa local zeta
function.}
\maketitle

\section{Introduction}

During the eighties several physical models using $p-$adic numbers were
proposed. Particularly \ various \ models of $p-$adic quantum mechanics \cite%
{K1}, \cite{rtvw}, \cite{VV}, \cite{VVZ}. \ As a consequence of this fact \
several new mathematical problems emerged, among them, the study of $p-$adic
pseudo-differential equations \cite{Koch1}, \cite{VVZ}. In this paper we
initiate the study of the decay of the solutions of wave-type
pseudo-differential equations over $p-$adic fields; these equations were
introduced by Kochubei \cite{Koch2} in connection with the problem of
characterizing the $p-$adic wave functions using pseudo-differential
operators. \ We show that the solutions of $p-$adic wave-type equations have
a decay similar to the solutions of \ classical generalized wave equations.

\smallskip

Let $K$ be a \ $p-$adic field, i.e. a finite extension of $\mathbb{Q}_{p}$.
Let $R_{K}$ be the valuation ring of $K$, $P_{K}$ the maximal ideal \ of \ $%
R_{K}$, and \ $\overline{K}=R_{K}/$ $P_{K}$ the residue field \ of $K$. Let $%
\pi $\ denote a fixed local parameter of $R_{K}$. The cardinality of $%
\overline{K}$ is denoted by $q$. For $z\in K$, $v(z)\in \mathbb{Z}\cup
\left\{ +\infty \right\} $ denotes the valuation of $z$, and $\left\vert
z\right\vert _{K}=q^{-v(z)}$. Let $\mathbb{S}(K^{n})$ denote the $\mathbb{C}$%
-vector space of Schwartz-Bruhat functions over $K^{n}$, the dual \ space $%
\mathbb{S}^{\prime }(K^{n})$ is the space of distributions over $K^{n}$. Let 
$\mathcal{F}$ denote the Fourier transform over\ $\mathbb{S}(K^{n+1})$. The
reader can consult \ any of the references \cite{GGP}, \cite{VVZ}, \cite{TA}
for \ an exposition of the theory of distributions over $p-$adic\ fields.

This article aims to study the \ following initial value problem: 
\begin{equation}
\left\{ 
\begin{array}{cc}
\left( Hu\right) \left( x,t\right) =0\text{,} & x\in K^{n}\text{, \ }t\in K
\\ 
&  \\ 
u\left( x,0\right) =f_{0}\left( x\right) , & 
\end{array}%
\right.  \label{i1}
\end{equation}%
where $n\geq 1$, $f_{0}\left( x\right) \in \mathbb{S}(K^{n})$, and 
\begin{equation*}
\left( H\Phi \right) \left( t,x\right) :=\mathcal{F}_{\left( \tau ,\xi
\right) \rightarrow \left( t,x\right) }^{-1}\left( \left\vert \tau -\phi
\left( \xi \right) \right\vert _{K}\mathcal{F}_{\left( t,x\right)
\rightarrow \left( \tau ,\xi \right) }\Phi \right) \text{, }\Phi \in 
\mathcal{\mathbb{S}}(K^{n+1})\text{,}
\end{equation*}
is a pseudo-differential operator with symbol $\left\vert \tau -\phi \left(
\xi \right) \right\vert _{K}$, where $\phi \left( \xi \right) $ is a
polynomial in $K\left[ \xi _{1},\ldots ,\xi _{n}\right] $ satisfying $\phi
(0)=0$. In the case in which $\phi \left( \xi \right) =a_{1}\xi
_{1}^{2}+\ldots +a_{n}\xi _{n}^{2}$, $H$ \ is called \ a \textit{Schr\"{o}%
dinger-type pseudo-differential operator}; \ this operator was introduced by
Kochubei in \cite{Koch2}. For $n=1$ the solution of (\ref{i1}) \ appears in
the formalism of $p-$adic quantum mechanics as the wave function for the
free particle \cite{VV}. The problem of characterizing \ the $p-$adic wave
functions as solutions of some pseudo-differential equation remains open. \ 

Let $\Psi \left( \cdot \right) $\ denote an additive character of $K$
trivial on $R_{K}$ \ but no on $P_{K}^{-1}$. By passing to the Fourier
transform in (\ref{i1}) \ one gets \ that 
\begin{equation*}
\left\vert \tau -\phi \left( \xi \right) \right\vert _{K}\mathcal{F}_{\left(
x,t\right) \rightarrow \left( \tau ,\xi \right) }u=0.
\end{equation*}%
Then any distribution \ of the form $\mathcal{F}^{-1}g$ with $g$ a
distribution supported on $\tau -\phi \left( \xi \right) =0$ is a solution.
By taking 
\begin{equation*}
g\left( \xi ,\tau \right) =\left( \mathcal{F}_{x\rightarrow \xi
}f_{0}\right) \delta \left( \tau -\phi \left( \xi \right) \right) ,
\end{equation*}%
where $\delta $ is the Dirac distribution, one gets 
\begin{equation}
u(x,t)=\int\limits_{K^{n}}\Psi \left( t\phi \left( \xi \right)
+\sum\limits_{i=1}^{n}x_{i}\xi _{i}\right) \left( \mathcal{F}_{x\rightarrow
\xi }f_{0}\right) \left( \xi \right) \left\vert d\xi \right\vert ,
\label{i3}
\end{equation}%
here $\left\vert d\xi \right\vert $\ is the Haar measure of $K^{n}$
normalized so that $vol(R_{K}^{n})=1$.

In this paper we show that\ the decay of $u(x,t)$ is completely similar to
the decay of the solution of the following initial value problem:

\begin{equation}
\left\{ 
\begin{array}{cc}
\frac{\partial u^{\text{arch}}\left( x,t\right) }{\partial t}=i\phi \left(
D\right) u^{\text{arch}}\left( x,t\right) \text{,} & x\in \mathbb{R}^{n}%
\text{, \ }t\in \mathbb{R} \\ 
&  \\ 
u^{\text{arch}}\left( x,0\right) =f_{0}\left( x\right) , & 
\end{array}%
\right.  \label{i4}
\end{equation}%
here $\phi \left( D\right) $ is a pseudo-differential operator having symbol 
$\phi \left( \xi \right) $. In this case 
\begin{equation}
u^{\text{arch}}\left( x,t\right) =\int\limits_{\mathbb{R}^{n}}\exp 2\pi
i\left( t\phi \left( \xi \right) +\sum\limits_{i=1}^{n}x_{i}\xi _{i}\right)
\left( \mathcal{F}_{x\rightarrow \xi }f_{0}\right) \left( \xi \right) d\xi
\label{i5}
\end{equation}%
is the solution of the initial value problem (\ref{i4}). If $\phi \left( \xi
\right) =\xi _{1}^{2}+\ldots +\xi _{n}^{2}$, i.e. $\phi \left( D\right) $ is
the Laplacian, $u^{\text{arch}}\left( x,t\right) $ satisfies 
\begin{equation}
\left\Vert u^{\text{arch}}\left( x,t\right) \right\Vert _{L^{\frac{2\left(
n+2\right) }{n}}}\leq c\left\Vert f_{0}\right\Vert _{_{L^{2}}},  \label{i6}
\end{equation}%
(see \cite{St}). If $n=1$ and $\phi \left( \xi \right) =\xi ^{3}$, $u^{\text{%
arch}}\left( x,t\right) $ satisfies 
\begin{equation}
\left\Vert u^{\text{arch}}\left( x,t\right) \right\Vert _{L^{8}}\leq
c\left\Vert f_{0}\right\Vert _{_{L^{2}}},  \label{i7}
\end{equation}%
(see \cite{KPV}). We show that $u\left( x,t\right) $ satisfies (\ref{i6}),
if $\phi \left( \xi \right) =\xi _{1}^{2}+\ldots +\xi _{n}^{2}$ (see \
Theorem \ref{cor1}), and that $u\left( x,t\right) $ satisfies (\ref{i7}), if 
$\phi \left( \xi \right) =\xi ^{3}$ (see \ Theorem \ref{cor2}). For more
general symbols we are able to describe \ the decay of $u\left( x,t\right) $
in $L^{\sigma }\left( K^{n+1}\right) $, however, in this case the index $%
\sigma $ is not optimal (see Theorem \ref{th6}). The proof is achieved by
adapting \ standard \ techniques \ in PDEs and by using number-theoretic
techniques for estimating exponential sums modulo $\pi ^{m}$. Indeed, like
in the classical case the estimation of the decay rate can be reduced to the
problem of estimating of the restriction of Fourier transforms to
non-degenerate hypersurfaces \cite{S2}; we solve this problem (see Theorems %
\ref{th4}, \ref{th5}) by reducing it to the estimation of exponential sums
modulo $\pi ^{m}$ (see Theorems \ref{th2a}, \ref{th2}). These exponential
sums are related to the Igusa zeta function for non-degenerate polynomials 
\cite{D1}, \cite{I1}, \cite{Z2}, \cite{Z3}. More precisely, by using Igusa's
method, the estimation of these exponential sums can be reduced to the
description of the poles of twisted local zeta functions \cite{D1}, \cite{Z2}%
, \cite{Z3}.

The restriction of Fourier transforms in $\mathbb{R}^{n}$ (see e.g. \cite[%
Chap. VIII]{S2}) was first posed and partially solved by Stein \cite{Fef}.
This problem \ have been intensively studied \ during the last thirty years 
\cite{B}, \cite{S2}, \cite{St}, \cite{T}. Recently Mockenhaupt and Tao have
studied the restriction problem in $\mathbb{F}_{q}^{n}$ \cite{MT}. In this
paper we \ initiate the study of the restriction problem in the
non-archimedean field setting.

The author \ thanks to the referee for his/her \ careful reading of this
paper.

\section{The Non-archimedean Principle of the Stationary Phase}

Given \ $f(x)\in K\left[ x\right] $, $x=\left( x_{1},\ldots ,x_{m}\right) $,
\ we denote by 
\begin{equation*}
C_{f}(K)=\left\{ z\in K^{m}\mid \frac{\partial f}{\partial x_{1}}\left(
z\right) =\cdots =\frac{\partial f}{\partial x_{m}}\left( z\right) =0\right\}
\end{equation*}%
the critical set of the mapping $f:K^{m}\rightarrow K$. If $f(x)\in R_{K}%
\left[ x\right] $, we denote by $\overline{f}\left( x\right) $ its reduction
modulo $\pi $, i.e. the polynomial obtained by reducing the coefficients of $%
f\left( x\right) $ modulo $\pi $.

Give a compact open set $A\subset K^{m}$, we set 
\begin{equation*}
E_{A}(z,f)=\int\limits_{A}\Psi \left( zf\left( x\right) \right) \left\vert
dx\right\vert \text{, }
\end{equation*}%
for $z\in K$, where $\left\vert dx\right\vert $ \ is the normalized Haar
measure of $K^{m}$. If $A=R_{K}^{m}$ we use the simplified notation $E(z,f)$
instead of $E_{A}(z,f)$. If $f(x)\in R_{K}\left[ x\right] $, then 
\begin{equation*}
E(z,f)=q^{-nm}\sum\limits_{x\text{ mod }\pi ^{n}}\Psi \left( zf\left(
x\right) \right) \text{;}
\end{equation*}%
thus $E(z,f)$ is a generalized Gaussian sum.

\begin{lemma}
\label{lemsp1}Let $f(x)\in R_{K}\left[ x\right] $, $x=\left( x_{1},\ldots
,x_{m}\right) $, be a non-constant polynomial. Let $A$ be the preimage of $%
\overline{\text{ }A}\subseteq \mathbb{F}_{q}^{m}$ under the canonical
homomorphism $R_{K}^{m}\rightarrow \left( R_{K}/P_{K}\right) ^{m}$. If $%
C_{f}(K)\cap A=\emptyset $, then there exists a constant $I(f,A)$ such that 
\begin{equation*}
E(z,f)=0\text{, \ \ for \ }\left| z\right| _{K}>q^{2I(f,A)+1}\text{.}
\end{equation*}
\end{lemma}

\begin{proof}
We define 
\begin{equation*}
I(f,a)=\min_{1\leq i\leq m}\left\{ v\left( \frac{\partial f}{\partial x_{i}}%
\left( a\right) \right) \right\} ,
\end{equation*}
for any $a\in A$, and 
\begin{equation*}
I\left( f,A\right) =\sup_{a\in A}\left\{ I(f,a)\right\} .
\end{equation*}
Since $A$ is compact and $C_{f}(K)\cap A=\emptyset $, $I\left( f,A\right)
<\infty $.

We denote by $a^{\ast }$\ an equivalence class of $R_{K}^{m}$ modulo $\left(
P_{K}^{I\left( f,A\right) +1}\right) ^{m}$, and by $a\in R_{K}^{m}$ a fixed
representative of $a^{\ast }$. By decomposing $\ A$ into equivalence classes
modulo $\left( P_{K}^{I\left( f,A\right) +1}\right) ^{m}$, one gets 
\begin{equation*}
E(z,f)=\sum_{a^{\ast }\subseteq A}q^{-m\left( I\left( f,A\right) +1\right)
}\int\limits_{R_{K}^{m}}\Psi \left( zf\left( a+\pi ^{I\left( f,A\right)
+1}x\right) \right) \left\vert dx\right\vert .
\end{equation*}%
Thus, it is sufficient to show that $\int\nolimits_{R_{K}^{m}}\Psi \left(
zf\left( a+\pi ^{I\left( f,A\right) +1}x\right) \right) \left\vert
dx\right\vert =0$ \ for \ $\left\vert z\right\vert _{K}>q^{2I(f,A)+1}$.

On the other hand, if $a=\left( a_{1},\ldots ,a_{m}\right) $, then 
\begin{equation*}
\frac{f\left( a+\pi ^{I\left( f,A\right) +1}x\right) -f\left( a\right) }{\pi
^{I\left( f,A\right) +1+\alpha _{0}}}
\end{equation*}%
equals 
\begin{equation*}
\sum\limits_{i=1}^{m}\pi ^{-\alpha _{0}}\frac{\partial f}{\partial x_{i}}%
\left( a\right) \left( x-a_{i}\right) +\pi ^{I\left( f,A\right) +1-\alpha
_{0}}\left( \text{higher order terms}\right) ,
\end{equation*}%
where 
\begin{equation*}
\alpha _{0}=\min_{i}\left\{ v\left( \frac{\partial f}{\partial x_{i}}\left(
a\right) \right) \right\} .
\end{equation*}%
Therefore 
\begin{equation}
f\left( a+\pi ^{I\left( f,A\right) +1}x\right) -f\left( a\right) =\pi
^{I\left( f,A\right) +1+\alpha _{0}}\widetilde{f}(x)  \label{sp2}
\end{equation}%
with $\widetilde{f}(x)\in R_{K}\left[ x\right] ,$ and since $C_{f}(K)\cap
A=\emptyset $, there exists an $i_{0}\in \left\{ 1,\ldots ,m\right\} $ such
that%
\begin{equation}
\overline{\frac{\partial \widetilde{f}}{\partial x_{i_{0}}}}\left( \overline{%
a}\right) \neq 0.  \label{sp3}
\end{equation}

We put $y=\Phi \left( x\right) =\left( \Phi _{1}\left( x\right) ,\ldots
,\Phi _{m}\left( x\right) \right) $ where 
\begin{equation*}
\Phi _{i}\left( x\right) =\left\{ 
\begin{array}{ccc}
\widetilde{f}(x) &  & i=i_{0} \\ 
x_{i} &  & i\neq i_{0}.%
\end{array}%
\right.
\end{equation*}%
Since $\Phi _{1}\left( x\right) ,\ldots ,\Phi _{m}\left( x\right) $ are
restricted power series and 
\begin{equation*}
\overline{J\left( \frac{\left( y_{1},\ldots ,y_{m}\right) }{\left(
x_{1},\ldots ,x_{m}\right) }\right) }=\overline{\frac{\partial \widetilde{f}%
}{\partial x_{i_{0}}}}\left( \overline{a}\right) \neq 0,
\end{equation*}%
the non-archimedean implicit \ function theorem implies that $y=\Phi \left(
x\right) $\ gives a measure-preserving map from $R_{K}^{m}$\ to $R_{K}^{m}$
(see \cite[Lemma 7.43]{I1}). Therefore 
\begin{eqnarray*}
\int\nolimits_{R_{K}^{m}}\Psi \left( zf\left( a+\pi ^{I\left( f,A\right)
+1}x\right) \right) \left\vert dx\right\vert &=& \\
\Psi \left( zf\left( a\right) \right) \int\nolimits_{R_{K}}\Psi \left( z\pi
^{I\left( f,A\right) +1+\alpha _{0}}y_{i_{0}}\right) \left\vert
dy_{i_{0}}\right\vert &=&0,
\end{eqnarray*}%
for $v\left( z\right) <-\left( I\left( f,A\right) +1+\alpha _{0}\right) $,
i.e. for $\left\vert z\right\vert _{K}>q^{I\left( f,A\right) +1+\alpha _{0}}$%
, and a fortiori 
\begin{equation*}
\int\nolimits_{R_{K}^{m}}\Psi \left( zf\left( a+\pi ^{I\left( f,A\right)
+1}x\right) \right) \left\vert dx\right\vert =0,
\end{equation*}%
for $\left\vert z\right\vert _{K}>q^{2I(f,A)+1}$ and any$\ a.$
\end{proof}

\begin{theorem}
\label{th1}Let $f(x)\in K\left[ x\right] $, $x=\left( x_{1},\ldots
,x_{m}\right) $, be a non-constant polynomial. Let $B\subset K^{m}$ be a
compact open set. If $C_{f}(K)\cap B=\emptyset $, then there exist \ a
constant $c\left( f,B\right) $ such that 
\begin{equation*}
E_{B}(z,f)={\LARGE 0}\text{, \ for }\left\vert z\right\vert _{K}\geq c\left(
f,B\right) .
\end{equation*}
\end{theorem}

\begin{proof}
By taking a covering $\cup _{i}\left( y_{i}+\left( \pi ^{\alpha
}R_{K}\right) ^{m}\right) $ of $B$, $E_{B}(z,f)$ can be expressed as \
linear combination of integrals of the form $E(z,f_{i})$ with $f_{i}(x)\in K%
\left[ x\right] $. After changing $z$ by $z\pi ^{\beta }$, we may suppose
that $f_{i}(x)\in R_{K}\left[ x\right] $. By applying Lemma \ref{lemsp1} we
get that $E(z,f_{i})=0$, \ \ for \ $\left| z\right| _{K}>c_{i}$. Therefore 
\begin{equation}
E_{B}(z,f)=0,\ \ \text{for}\ \left| z\right| _{K}>\max_{i}c_{i}.  \label{sp5}
\end{equation}
\end{proof}

We note that the previous result implies that 
\begin{equation*}
E_{B}(z,f)=O(\left| z\right| _{K}^{-M}),
\end{equation*}
for any $M\geq 0$. This is the standard form of the principle of the
stationary phase.\ 

\section{Local Zeta Functions and Exponential Sums}

In this section we review some results about exponential sums and Newton
polyhedra that will be used in the next section. For $x\in K$ we denote by $%
ac(x)=x\pi ^{-v(x)}$ its angular component. Let $f(x)\in R_{K}[x]$, $%
x=(x_{1},\ldots ,x_{m})$ be a non-constant polynomial, and $\chi
:R_{K}^{\times }\longrightarrow \mathbb{C}^{\times }$ a character of $%
R_{K}^{\times }$, the group of units of $R_{K}$. We formally put $\chi (0)=0$%
. To these data one associates the Igusa local zeta function,

\begin{equation*}
Z(s,f,\chi )=\int_{R_{K}^{n}}\chi (acf(x))|f(x)|_{K}^{s}\mid dx\mid
,\,\,\,s\in \mathbb{C},\,\,
\end{equation*}%
for $Re(s)>0$, where $\mid dx\mid $ denotes the normalized Haar measure of $%
K^{n}$. The Igusa local zeta function admits a meromorphic continuation to
the complex plane as a rational function of $q^{-s}$. Furthermore, \ it is
related to the number of solutions of polynomial congruences modulo $\pi
^{m} $\ and exponential sums modulo $\pi ^{m}$ \cite{D0}, \cite{I1}.

\subsection{Exponential Sums Associated with Non-degenerate Polynomials}

We set $\mathbb{R}_{+}=\{x\in \mathbb{R}\,\,\mid \,\,x\geqq 0\}$. Let $%
f(x)=\sum_{l}a_{l}x^{l}\in K[x]$, $x=(x_{1},\ldots ,x_{m})$ be a \
non-constant polynomial satisfying $f(0)=0$. The set $supp(f)=\{l\in \mathbb{%
N}^{m}\mid \,a_{l}\neq 0\}$ is called the support of $f$. The Newton
polyhedron $\Gamma (f)$ of $f$ is defined as the convex hull in $\mathbb{R}%
_{+}^{m}$ of the set

\begin{equation*}
\bigcup_{l\in supp(f)}\left( l+\mathbb{R}_{+}^{m}\right) .
\end{equation*}%
We denote by $\left\langle \cdot ,\cdot \right\rangle $ the usual inner
product of $\mathbb{R}^{m}$, and identify $\mathbb{R}^{m}$ with its dual by
means of it. We set 
\begin{equation*}
\left\langle a_{\gamma },x\right\rangle =m(a_{\gamma }),
\end{equation*}%
for the equation of the supporting hyperplane of a facet $\gamma $ (i.e. a
face of codimension $1$ of $\Gamma (f)$) with perpendicular vector $%
a_{\gamma }=(a_{1},\ldots ,a_{n})\in \mathbb{N}^{n}\smallsetminus \{0\}$,
and $\sigma \left( a_{\gamma }\right) :=\sum_{i}a_{i}$.

\begin{definition}
A polynomial $f(x)\in K[x]$ is called \textit{\ non-degenerate with respect
to its Newton polyhedron } $\Gamma (f)$, if it satisfies the following two
properties: (i) $C_{f}\left( K\right) =\left\{ 0\right\} $ $\subset $ $K^{n}$%
; (ii) for every proper face $\gamma \subset \Gamma (f)$, the critical set $%
C_{f_{\gamma }}\left( K\right) $ of $f_{\gamma }(x):=\sum_{i\in \gamma
}a_{i}x^{i}$ satisfies $C_{f_{\gamma }}\left( K\right) $ $\cap
(K\smallsetminus \{0\})^{m}=\emptyset .$
\end{definition}

We note that the above definition is not standard because it requires that
the origin be an isolated critical point (see e.g. \cite{D1}, \cite{D-H}, 
\cite{Z3}). The condition (ii) can be replaced by 
\begin{equation}
\left\{ x\in K^{m}\mid f_{\gamma }(x)=0\right\} \cap C_{f_{\gamma }}\left(
K\right) \cap (K\smallsetminus \{0\})^{m}=\emptyset .  \label{lz1}
\end{equation}

If $K$ has characteristic $p>0$, by using Euler's identity, it can be
verified that condition (ii) in the above definition is equivalent to (\ref%
{lz1}), if $p$ does not divide the $m(a_{\gamma })\neq 0$, for any facet $%
\gamma $.

In \cite{Z3} the author showed that if $f$ is non-degenerate with respect $%
\Gamma (f)$,\ then the \ poles of $\left( 1-q^{-1-s}\right) Z(s,f,\chi _{%
\text{triv}})$ and $Z(s,f,\chi )$, $\chi \neq \chi _{\text{triv}}$, have the
form 
\begin{equation*}
s=-\frac{\sigma \left( a_{\gamma }\right) }{m(a_{\gamma })}+\frac{2\pi i}{%
log\,q}\frac{k}{m(a_{\gamma })}\text{,}\,\,k\in \mathbb{Z}\text{,}
\end{equation*}%
for some facet $\gamma $ of $\Gamma (f)$ with perpendicular $a_{\gamma }$,
and $m(a_{\gamma })\neq 0$ (see \cite[Theorem A, and Lemma 4.4]{Z3}).
Furthermore, if $\chi \neq \chi _{\text{triv}}$ and the order of $\chi $
does not divide any $m(a_{\gamma })\neq 0$, where $\gamma $ is a facet of $%
\Gamma (f)$, then $Z(s,f,\chi )$ is a polynomial in $q^{-s}$, and its degree
is bounded by a constant independent of $\chi $ (see \cite[Theorem B]{Z3}).
These two results imply that for $\mid z\mid _{K}$ big enough $E(z,f)$ is a
finite $\mathbb{C}$-linear combination of functions of the form 
\begin{equation*}
\chi (ac(z))\mid z\mid _{K}^{\lambda }(log_{q}(|z|_{K}))^{\gamma },
\end{equation*}%
with coefficients independent of $z$, and with $\lambda \in \mathbb{C}$ a
pole of 
\begin{equation*}
(1-q^{-1-s})Z(s,f,\chi _{\text{triv}})\text{ or of }Z(s,f,\chi ),\chi \neq
\chi _{\text{triv}},
\end{equation*}%
and $\gamma \in \mathbb{N}$, $\gamma \leqq $(multiplicity of pole $\lambda $%
) $-1$ (see \cite[Corollary 1.4.5]{D0}). Moreover all poles $\lambda $
appear effectively in this linear combination. Therefore 
\begin{equation}
|E(z,f)|\leqq C\mid z\mid _{K}^{-\beta _{f}+\epsilon },  \label{bound1}
\end{equation}%
with $\epsilon >0$, and 
\begin{equation*}
\beta _{f}:=\min_{\tau }\{\frac{\sigma \left( a_{\tau }\right) }{m(a_{\tau })%
}\},
\end{equation*}%
where $\tau $ runs through all facets of $\Gamma (f)$ satisfying $m(a_{\tau
})\neq 0$. The point 
\begin{equation*}
T_{0}=(\beta _{f}^{-1},...,\beta _{f}^{-1})\in \mathbb{Q}^{m}
\end{equation*}%
is the intersection point of the boundary of the Newton polyhedron $\Gamma
(f)$ with the diagonal $\Delta =\{(t,\ldots ,t)\;\mid \,t\in \mathbb{R}\}$ $%
\subset \mathbb{R}^{m}$. By combining estimation (\ref{bound1}) and \
Theorem \ref{th1}, we obtain the following result.

\begin{theorem}
\label{th2a}Let $f(x)\in K[x]$ be \textit{\ non-degenerate with respect to
its Newton polyhedron } $\Gamma (f)$. Let $B\subset K^{m}$ a compact open
subset. Then 
\begin{equation*}
|E_{B}(z,f)|\leqq C\mid z\mid _{K}^{-\beta _{f}+\epsilon },
\end{equation*}%
for any $\epsilon >0$.
\end{theorem}

We have to mention that the previous result is known by the experts, however
the author did not find a suitable reference for the purposes of this
article. If $K$ has characteristic $p>0$, the previous result is valid if $p$
does not divide the $m(a_{\tau })\neq 0$ \cite[Corollary 6.1]{Z3}. \ \ 

\subsection{Exponential Sums Associated with Quasi-homogeneous Polynomials}

\begin{definition}
Let $f(x)\in K[x],$ $x=\left( x_{1},\ldots ,x_{m}\right) $ be a \
non-constant polynomial satisfying $f(0)=0$. The polynomial $f(x)$ is called
quasi-homogeneous of degree $d$ with respect $\alpha =\left( \alpha
_{1},\ldots ,\alpha _{m}\right) \in \left( \mathbb{N}\setminus \left\{
0\right\} \right) ^{m}$, if it satisfies 
\begin{equation*}
f(\lambda ^{\alpha _{1}}x_{1},\ldots ,\lambda ^{\alpha _{m}}x_{m})=\lambda
^{d}f\left( x\right) ,\text{ for every }\lambda \in K.
\end{equation*}%
In addition, if \ $C_{f}(K)$ is the origin of $K^{m}$, then $f(x)$ is called
a non-degenerate quasi-homogeneous polynomial.
\end{definition}

The \ non-degenerate quasi-homogeneous polynomials are a subset \ of the
non-degenerate polynomials with respect to the Newton polyhedron. For these
type of polynomials the bound (\ref{bound1}) can be improved: 
\begin{equation}
\left\vert E(z,f)\right\vert \leq C\left\vert z\right\vert _{K}^{-\beta
_{f}},  \label{bound2}
\end{equation}%
where $\beta _{f}=\frac{1}{d}\sum_{i=1}^{m}\alpha _{i}$. By using the
techniques exposed in \cite[Theorem 3.5]{Z2}, and \cite[Lemma 2.4]{Z3}
follow that the \ poles of $\left( 1-q^{-1-s}\right) Z(s,f,\chi _{\text{triv}%
})$ and $Z(s,f,\chi )$, $\chi \neq \chi _{\text{triv}}$, have the form 
\begin{equation*}
s=-\frac{\sigma \left( \alpha \right) }{d}+\frac{2\pi i}{log\,q}\frac{k}{d}%
\text{,}\,\,k\in \mathbb{Z}\text{.}
\end{equation*}%
Then by using the same reasoning as before, we obtain (\ref{bound2}). This
estimate and Theorem \ref{th1} imply the following result.

\begin{theorem}
\label{th2}Let $f(x)\in K[x],$ $x=\left( x_{1},...,x_{m}\right) $ be a \
non-degenerate quasi-homogeneous polynomial of degree $d$ with respect to $%
\alpha =\left( \alpha _{1},\ldots ,\alpha _{m}\right) $. Let $B$ $\subset
K^{m}$ be a compact open set. Then 
\begin{equation*}
\left\vert E_{B}\left( z,f\right) \right\vert \leq C\left\vert z\right\vert
_{K}^{-\beta _{f}}.
\end{equation*}
\end{theorem}

If $K$ has characteristic $p>0$, the above result is valid, if $p$ does not
divide $\sigma \left( \alpha \right) $.

\section{Fourier Transform of Measures Supported on Hypersurfaces}

Let $Y$ \ be a closed smooth submanifold of $K^{n}$ of dimension $n-1$. \ If 
\begin{equation}
I=\left\{ i_{1},\ldots ,i_{n-1}\right\} \text{ with }\ i_{1}<i_{2}<\ldots
<i_{n-1}  \label{p1}
\end{equation}%
is a subset of $\left\{ 1,\ldots ,n\right\} $ we denote by $\omega _{Y_{I}}$
\ the differential form induced on $Y$ by $dx_{i_{1}}\wedge dx_{i_{2}}\wedge
\ldots \wedge dx_{i_{n-1}}$, \ and by $d\sigma _{Y_{I}}$ the corresponding
measure on $Y$. \ The canonical measure of $Y$ is defined as 
\begin{equation*}
d\sigma _{Y}=\sup_{I}\text{ }\left\{ d\sigma _{Y_{I}}\right\}
\end{equation*}%
where \ $I$ runs through all the subsets of form (\ref{p1}). Given $S$ \ a
compact open subset of $K^{n}$ with characteristic function $\Theta _{S}$,
we define \ $d\mu _{Y,S}=d\mu _{Y}=\Theta _{S}d\sigma _{Y}$. The \ canonical
measure $d\mu _{Y}$ was introduced by Serre in \cite{Ser}. The \ Fourier \
transform of $d\mu _{Y}$\ is defined as 
\begin{equation*}
\widehat{d\mu _{Y}\left( \xi \right) }=\int\limits_{Y}\Psi \left( -\left[
x,\xi \right] \right) d\mu _{Y}\left( x\right) ,
\end{equation*}%
where $\left[ x,y\right] :=\sum_{i=1}^{n}x_{i}y_{i}$, with \ $x$, $y\in
K^{n} $. The analysis of the decay of $\widehat{\left\vert d\mu _{Y}\left(
\xi \right) \right\vert }$ as $\left\Vert \xi \right\Vert
_{K}:=\max_{i}\left\{ \left\vert \xi _{i}\right\vert _{K}\right\} $
approaches infinity plays a central role in this paper. \ This analysis can
be simplified taking into account the following facts. Any compact open set
of $K^{n}$ is a finite union of classes modulo $\pi ^{e}$, by taking $e$ big
enough, and taking into account that \ $Y\cap $ $y+(\pi ^{e}R_{K})^{n}$ is \
a hypersurface of the form 
\begin{equation*}
\left\{ x\in y+(\pi ^{e}R_{K})^{n}\mid x_{n}=\phi \left( x_{1},\ldots
,x_{n-1}\right) \right\}
\end{equation*}%
with $\phi $\ an analytic function satisfying 
\begin{equation}
\phi \left( 0\right) =\frac{\partial \phi }{\partial x_{1}}\left( 0\right)
=\ldots =\frac{\partial \phi }{\partial x_{n-1}}\left( 0\right) =0,
\label{grad}
\end{equation}%
(see \cite[page 147]{Ser}), we may assume that $Y$ is a hypersurface \ of \
the form \ $x_{n}-\phi \left( x_{1},\ldots ,x_{n-1}\right) =0$, $\ $\ with $%
\phi $ satisfying (\ref{grad}). In this case \ $d\sigma _{Y}\left( x\right)
=\left\vert dx_{1}\right\vert \ldots \left\vert dx_{n-1}\right\vert $, the
normalized Haar measure of $K^{n-1}$.

Finally we want to mention that if $X=\left\{ x\in K^{n}\mid f(x)=0\right\} $
is a hypersurface then 
\begin{equation*}
\frac{dx_{1}\ldots dx_{n-1}}{\left\vert \frac{\partial f}{\partial x_{n}}%
\right\vert _{K}}
\end{equation*}%
is a measure on a neighborhood of $X$ provided that $\left\vert \frac{%
\partial f}{\partial x_{n}}\right\vert _{K}\neq 0$ (see \cite[Sect. 7.6]{I1}%
). This measure is not \ intrinsic to $X$, but if $S$ is small enough, it
coincides with $d\mu _{X}=\Theta _{S}d\sigma _{X}$ for a polynomial of type $%
f(x)=x_{n}-\phi \left( x_{1},\ldots ,x_{n-1}\right) $. The Serre measure
allow us to \ define $\widehat{d\mu _{Y}\left( \xi \right) }$\ intrinsically
for an arbitrary submanifold $Y$.

\begin{theorem}
\label{th4}Let $\phi \left( x\right) \in R_{K}\left[ x\right] $\ , $x=\left(
x_{1},\ldots ,x_{n-1}\right) $, be a non-constant \ polynomial such that $%
C_{\phi }(K)=\left\{ 0\right\} \subset K^{n-1}$. Let $d_{j}(\phi )$ be the
degree of $\phi $ with respect the variable $x_{j}$, and let $\beta _{\phi
}:=\max_{j}d_{j}(\phi )$. Let $\Theta _{S}$ be the characteristic function
of a compact open set $S$, let 
\begin{equation*}
Y=\left\{ x\in K^{n}\mid x_{n}=\phi \left( x_{1},\ldots ,x_{n-1}\right)
\right\} ,
\end{equation*}%
and let $d\mu _{Y}=\Theta _{S}d\sigma _{Y}$. Then 
\begin{equation}
\left\vert \widehat{d\mu _{Y}\left( \xi \right) }\right\vert \leq
C\left\Vert \xi \right\Vert _{K}^{-\beta },  \label{ft1}
\end{equation}%
for $\ 0\leq \beta \leq \beta _{\phi }-\epsilon $, with $\epsilon >0$.
\end{theorem}

\begin{proof}
By passing to a sufficiently fine covering we may suppose that\ 
\begin{equation*}
\widehat{d\mu _{Y}\left( \xi \right) }=\int\limits_{\left( z_{0}+\pi
^{e_{0}}R_{K}\right) ^{n-1}}\Psi \left( -\xi _{n}\phi \left( x\right) -\left[
x,\xi ^{\prime }\right] \right) \left\vert dx\right\vert .
\end{equation*}%
By applying Theorem 6.1 \ of \cite{Cluc}, \ we have

\begin{equation*}
\left\vert \widehat{d\mu _{Y}\left( \xi \right) }\right\vert \leq C\left(
\log _{q}\left\Vert \xi \right\Vert _{K}\right) ^{n-1}\left\Vert \xi
\right\Vert _{K}^{-\beta _{\phi }},
\end{equation*}%
and then 
\begin{equation*}
\left\vert \widehat{d\mu _{Y}\left( \xi \right) }\right\vert \leq
C\left\Vert \xi \right\Vert _{K}^{-\beta }\text{, for }0\leq \beta \leq
\beta _{\phi }-\epsilon \text{, }\epsilon >0.
\end{equation*}%
It is important to mention that Cluckers' \ Theorem 6.1 is established only
for $\mathbb{Q}_{p}$, however this result is valid for any $p$-adic field.
Indeed, the proof of this result \ is based on a result of Chubarikov \cite[%
Lemma 3]{Chuba} whose proof uses inductively an estimation for
one-dimensional exponential sums due to I. M.\ Vinogradov (see e.g. \cite[%
Theorem 2.1]{A-C-K}). The proof of this last estimation as given in \cite%
{A-C-K} can be adapted to the case of $p$-adic fields easily using the
notion of dilation as in \cite{Z2}.
\end{proof}

The Cluckers' result does not give an optimal decay rate, and then $\beta
_{\phi }$\ is not optimal (see also \cite{Z4}).

\begin{remark}
\label{Rem2} If $\phi \left( x\right)
=\dsum\nolimits_{i=1}^{n-1}a_{i}x_{i}^{2}$, then the phase of $\widehat{d\mu
_{Y}\left( \xi \right) }$ around any critical point has the form $%
\dsum\nolimits_{i=1}^{n-1}a_{i}^{\prime }x_{i}^{2}$. By using Theorem \ref%
{th2}, \ one verifies that the decay rate around the point is $\frac{n-1}{2}$%
, therefore Theorem \ref{th4} \ holds \ for $0\leq \beta \leq $ $\frac{n-1}{2%
}:=\beta _{\phi }$. If $n=1$ and $\phi \left( x\right) =x^{d}$, $d>1$, the
phase of $\widehat{d\mu _{Y}\left( \xi \right) }$ around a critical point
can take the form $x^{f}p\left( x\right) $, $2\leq f\leq d$, $p\left(
x\right) \neq 0$ locally. By using the fact the real parts of the possible
poles of the corresponding local zeta functions have the form $\frac{-1}{f}$%
, $2\leq f\leq d$, and \ \ Theorem 8.4.2 in \cite{I1}, one verifies that
Theorem \ref{th4} \ holds \ for $0\leq \beta \leq $ $\frac{1}{d}:=$ $\beta
_{\phi }$.
\end{remark}

In the case of real numbers the results described in the previous remark are
well-known (see e.g. \cite{S2}).

\subsection{Restriction of the Fourier Transform to Non-degenerate
Hypersurfaces}

Let $X$ be a submanifold of $K^{n}$ with $d\sigma _{X}$\ its canonical
measure. We set $d\mu _{Y,S}=\Theta _{S}d\sigma _{Y}$, where $\Theta _{S}$\
is the characteristic function of a compact open set $S$ in $K^{n}$. We say
\ that the $L^{\rho }$\textit{\ restriction property} \ is valid for $X$ if
there exists a $\tau \left( \rho \right) $\ so that 
\begin{equation*}
\left( \int\limits_{X}\left\vert \mathcal{F}g\left( \xi \right) \right\vert
_{K}^{\tau }d\mu _{X,S}\left( \xi \right) \right) ^{\frac{1}{\tau }}\leq
C_{\tau ,\rho }\left( S\right) \left\Vert g\right\Vert _{L^{\rho }}
\end{equation*}%
holds for each \ $g\in \mathbb{S}\left( K^{n}\right) $ $\ $and any \ compact
open set $S$ of $K^{n}$. \ 

The restriction problem in $\mathbb{R}^{n}$ (see e.g. \cite[Chap. VIII]{S2})
was first posed and partially solved by Stein \cite{Fef}. This problem \
have been intensively studied \ during the last thirty years \cite{B}, \cite%
{S2}, \cite{St}, \cite{T}. Recently Mockenhaupt and Tao have studied the
restriction problem in $\mathbb{F}_{q}^{n}$ \cite{MT}. In this paper we
study the restriction problem in the non-archimedean field setting. More
precisely, in the case in which $X$ is a non-degenerate hypersurface and $%
\tau =2$. The proof of \ the restriction property in the non-archimedean
case uses the Lemma of interpolation of operators (see e.g. \cite[Chap. IX]%
{S2}) and the estimates for oscillatory integrals \ obtained in the previous
section. The interpolation Lemma given in \cite[Chap. IX]{S2} is valid in
the non-archimedean case. For the sake of completeness we rewrite this lemma
here.

Let $\left\{ U^{z}\right\} $ be a family of operators on the strip $a\leq 
\func{Re}(z)\leq b$ defined by 
\begin{equation*}
\left( U^{z}g\right) \left( x\right) =\int\limits_{K^{n}}\mathfrak{K}%
_{z}\left( x,y\right) g\left( y\right) \left| dy\right| ,
\end{equation*}
where the kernels $\mathfrak{K}_{z}\left( x,y\right) $ have a fixed compact
support and are uniformly bounded for $(x,y)\in K^{n}\times K^{n}$ and $%
a\leq \func{Re}(z)\leq b$. We also assume that for each $(x,y)$, the
function $\mathfrak{K}_{z}\left( x,y\right) $ is analytic in $a<\func{Re}%
(z)<b$ and is continuous in the closure $a\leq \func{Re}(z)\leq b$, and that 
\begin{equation*}
\left\{ 
\begin{array}{cc}
\left\| U^{z}g\right\| _{L^{\tau _{0}}}\leq M_{0}\left\| g\right\| _{L^{\rho
_{0}}}\text{,} & \text{when }\func{Re}(z)=a\text{,} \\ 
&  \\ 
\left\| U^{z}g\right\| _{L^{\tau _{1}}}\leq M_{1}\left\| g\right\| _{L^{\rho
_{1}}}\text{,} & \text{when }\func{Re}(z)=b\text{;}%
\end{array}
\right.
\end{equation*}
here $\left( \tau _{i},\rho _{i}\right) $ are two pairs of given exponents
with \ $1\leq \tau _{i},\rho _{i}\leq \infty $.

\begin{lemma}[{Interpolation Lemma \protect\cite[Chap. IX]{S2}}]
Under the above \ hypo\-theses, 
\begin{equation*}
\left\Vert U^{a(1-\theta )+b\theta }g\right\Vert _{L^{\tau }}\leq
M_{0}^{1-\theta }M_{1}^{\theta }\left\Vert g\right\Vert _{L^{\rho }}
\end{equation*}%
where $\ 0\leq \theta \leq 1$, $\frac{1}{\tau }=\frac{\left( 1-\theta
\right) }{\tau _{0}}+\frac{\theta }{\tau _{1}}$, and $\frac{1}{\rho }=\frac{%
\left( 1-\theta \right) }{\rho _{0}}+\frac{\theta }{\rho _{1}}$.
\end{lemma}

\begin{theorem}
\label{th5}Let $\phi \left( x\right) \in R_{K}\left[ x\right] $\ , $x=\left(
x_{1},\ldots ,x_{n-1}\right) $, be a non-constant \ polynomial such that $%
C_{\phi }(K)=\left\{ 0\right\} \subset K^{n-1}$. Let 
\begin{equation*}
Y=\left\{ x\in K^{n}\mid x_{n}=\phi \left( x_{1},\ldots ,x_{n-1}\right)
\right\}
\end{equation*}%
with the measure $d\mu _{Y,S}=\Theta _{S}d\sigma _{Y}$, where $\Theta _{S}$\
is the characteristic function of a compact open subset $S$ of $K^{n}$. Then 
\begin{equation}
\left( \int\limits_{Y}\left\vert \mathcal{F}g\left( \xi \right) \right\vert
_{K}^{2}d\mu _{Y}\left( \xi \right) \right) ^{\frac{1}{2}}\leq C\left(
Y\right) \left\Vert g\right\Vert _{L^{\rho }},  \label{ret}
\end{equation}%
holds for each $1\leq \rho <\frac{2\left( 1+\beta _{\phi }\right) }{2+\beta
_{\phi }}$.
\end{theorem}

\begin{proof}
We first note that 
\begin{eqnarray}
\int\limits_{Y}\left\vert \mathcal{F}g\left( \xi \right) \right\vert
_{K}^{2}d\mu _{Y,S}\left( \xi \right) &=&\int\limits_{Y}\mathcal{F}g\left(
\xi \right) \overline{\mathcal{F}g\left( \xi \right) }d\mu _{Y,S}\left( \xi
\right)  \notag \\
&=&\int\limits_{K^{n}}\left( Tg\right) \left( x\right) \overline{g\left(
x\right) }\left\vert dx\right\vert  \label{rf2}
\end{eqnarray}%
where $\left( Tg\right) \left( x\right) =\left( g\ast \mathfrak{K}\right)
\left( x\right) $\ with 
\begin{equation*}
\mathfrak{K}\left( x\right) =\int\limits_{Y}\Psi \left( \left[ x,\xi \right]
\right) d\mu _{Y,S}\left( \xi \right) =\widehat{d\mu _{Y,S}\left( -x\right) }%
.
\end{equation*}%
The theorem follows from (\ref{rf2}) by H\"{o}lder's inequality \ if we show
that 
\begin{equation*}
\left\Vert T(g)\right\Vert _{L^{\rho _{0}^{\prime }}}\leq C\left\Vert
g\right\Vert _{L^{\rho _{0}}}
\end{equation*}%
where $\rho _{0}^{\prime }$ is the dual exponent of $\rho _{0}$. Now we
define $\mathfrak{K}_{z}\left( x\right) $\ as equal to 
\begin{equation*}
\gamma \left( z\right) \int\limits_{K^{n}}\Psi \left( \left[ x,\xi \right]
\right) \left\vert \xi _{n}-\phi \left( \xi ^{\prime }\right) \right\vert
_{K}^{-1+z}\eta \left( \xi _{n}-\phi \left( \xi ^{\prime }\right) \right)
\Theta _{S}\left( \xi ^{\prime },\phi \left( \xi ^{\prime }\right) \right)
\left\vert d\xi \right\vert ,
\end{equation*}%
where $\gamma \left( z\right) =\left( \frac{1-q^{-z}}{1-q^{-1}}\right) $, $%
\xi ^{\prime }=\left( \xi _{1},\ldots ,\xi _{n-1}\right) $, $\eta \left( \xi
\right) $\ is the characteristic function of the ball $P_{K}^{e_{0}}$, $%
e_{0}\geq 1$, and $\func{Re}(z)>0$. \ By making $y=\xi _{n}-\phi \left( \xi
^{\prime }\right) $ in the above integral we obtain 
\begin{equation*}
\mathfrak{K}_{z}\left( x\right) =\zeta _{z}\left( x_{n}\right) \mathfrak{K}%
(x)
\end{equation*}%
with 
\begin{equation*}
\zeta _{z}\left( x_{n}\right) =\gamma \left( z\right) \int\limits_{K}\Psi
\left( x_{n}y\right) \left\vert y\right\vert _{K}^{-1+z}\eta \left( y\right)
\left\vert dy\right\vert ,\text{ }\func{Re}(z)>0\text{.}
\end{equation*}

On the other hand, 
\begin{equation*}
\zeta _{z}\left( x_{n}\right) =\left\{ 
\begin{array}{cc}
q^{-e_{0}z}, & \text{if \ }\left\vert x_{n}\right\vert _{K}\leq q^{e_{0}};
\\ 
\left( \frac{1-q^{z-1}}{1-q^{-1}}\right) \left\vert x_{n}\right\vert
_{K}^{-z}, & \text{if \ }\left\vert x_{n}\right\vert _{K}>q^{e_{0}},%
\end{array}%
\right.
\end{equation*}%
(for a similar calculation the reader can see \cite[page 54]{Vla}), then $%
\zeta _{z}\left( x_{n}\right) $\ has an analytic continuation \ to the
complex plane \ as an entire function; also $\zeta _{0}\left( x_{n}\right)
=1 $, and $\left\vert \zeta _{z}\left( x_{n}\right) \right\vert \leq
c\left\vert x_{n}\right\vert _{K}^{-\func{Re}\left( z\right) }$ where $%
\left\vert x_{n}\right\vert _{K}\geq q^{e_{0}}$. Therefore $\zeta _{z}\left(
x_{n}\right) $\ has an analytic continuation \ to an entire function
satisfying the following properties:

(i) $\mathfrak{K}_{0}\left( x\right) =\mathfrak{K}\left( x\right) $,

(ii) $\left\vert \mathfrak{K}_{-\beta +i\gamma }\left( x\right) \right\vert
\leq C$, for every $x\in K^{n}$, $\gamma \in \mathbb{R}$, and $0\leq \beta
\leq \beta _{\phi }-\epsilon $, $\epsilon >0$,

(iii) $\left\vert \mathcal{F}\mathfrak{K}_{1+i\gamma }\left( \xi \right)
\right\vert \leq C$, for $\xi \in K^{n}$, and $\ $\ $\gamma \in \mathbb{R}$.

In fact (ii) follows from Theorem \ref{th4} , and (iii) is an immediate
consequence of the definition of $\mathfrak{K}_{z}\left( x\right) $.

Now we consider the analytic family of operators $T_{z}(g)=\left( g\ast 
\mathfrak{K}_{z}\right) \left( x\right) $. From (ii) one has 
\begin{equation*}
\left\Vert T_{-\beta +i\gamma }\left( g\right) \right\Vert _{L^{\infty
}}\leq C\left\Vert g\right\Vert _{L^{1}}\text{, }
\end{equation*}%
for $0\leq \beta \leq \beta _{\phi }-\epsilon $, $\epsilon >0$, and $\gamma
\in \mathbb{R}$, and from (iii) and Plancherel's Theorem one gets 
\begin{equation*}
\left\Vert T_{1+i\gamma }\left( g\right) \right\Vert _{L^{2}}\leq
C\left\Vert g\right\Vert _{L^{2}}\text{, }
\end{equation*}%
for $\gamma \in \mathbb{R}$. By applying the Interpolation Lemma with 
\begin{equation*}
\theta =\frac{\beta }{1+\beta },
\end{equation*}%
we obtain 
\begin{equation*}
\left\Vert T_{0}\left( g\right) \right\Vert _{L^{\rho ^{\prime }}}\leq
C\left\Vert g\right\Vert _{L^{\rho }}\text{, }
\end{equation*}%
with $\rho ^{\prime }$ the dual exponent of $\rho =\frac{2\left( 1+\beta
\right) }{2+\beta }$, and $0\leq \beta \leq \beta _{\phi }-\epsilon $, $%
\epsilon >0$. Therefore the previous estimate for $\left\Vert T_{0}\left(
g\right) \right\Vert _{L^{\rho ^{\prime }}}$ is valid for $1\leq \rho \leq 
\frac{2\left( 1+\beta _{\phi }-\epsilon \right) }{2+\beta _{\phi }-\epsilon }
$.
\end{proof}

Our proof of Theorem \ref{th5} is strongly influenced by \ Stein's proof \
for the restriction problem in the case of a smooth hypersurface in $\mathbb{%
R}^{n}$ with non-zero Gaussian curvature \cite{S1a}.

\section{Asymptotic Decay of Solutions of Wave-type Equations}

Like in the classical case \cite{St}, the decay of the solutions of
wave-type pseudo-differential equations can be deduced \ from the
restriction theorem proved in the previous section, taking into account that
the following two problems are completely equivalent if $\frac{1}{\rho }+%
\frac{1}{\sigma }=1$:

\begin{problem}
For which values of $\rho $, $1\leq \rho <2$, is it true that $f\in L^{\rho
}(K^{n})$ implies that $\mathcal{F}f$\ has a well-defined restriction \ to $%
Y $ in $L^{2}(d\mu _{Y,S})$ with 
\begin{equation*}
\left( \int\limits_{Y}\left\vert \mathcal{F}f\right\vert ^{2}d\mu
_{Y,S}\right) ^{\frac{1}{2}}\leq C_{\rho }\left\Vert f\right\Vert _{L^{\rho
}}?
\end{equation*}
\end{problem}

\begin{problem}
For which values of $\sigma $, $2<\sigma \leq \infty $, is it true that the
distribution $gd\mu _{Y,S}$ for each $g\in L^{2}(d\mu _{Y,S})$ \ has Fourier
transform in $L^{\sigma }(K^{n})$ with 
\begin{equation*}
\left\Vert \mathcal{F}\left( gd\mu _{Y,S}\right) \right\Vert _{L^{\sigma
}}\leq C_{\sigma }\left( \int\limits_{Y}\left\vert g\right\vert ^{2}d\mu
_{Y,S}\right) ^{\frac{1}{2}}?
\end{equation*}
\end{problem}

\subsection{Wave-type Equations with Non-degenerate Symbols}

\begin{theorem}[Main Result]
\label{th6}Let $\phi \left( \xi \right) \in R_{K}\left[ \xi \right] $\ , $%
\xi =\left( \xi _{1},\ldots ,\xi _{n}\right) $, be a non-constant \
polynomial such that $C_{\phi }(K)=\left\{ 0\right\} \subset K^{n}$. Let 
\begin{equation*}
\left( H\Phi \right) \left( t,x\right) =\mathcal{F}_{\left( \tau ,\xi
\right) \rightarrow \left( t,x\right) }^{-1}\left( \left\vert \tau -\phi
\left( \xi \right) \right\vert _{K}\mathcal{F}_{\left( t,x\right)
\rightarrow \left( \tau ,\xi \right) }\Phi \right) \text{, }\Phi \in 
\mathcal{\mathbb{S}}(K^{n+1})\text{,}
\end{equation*}%
be a pseudo-differential operator with symbol $\left\vert \tau -\phi \left(
\xi \right) \right\vert _{K}$. \ Let $u(x,t)$ be the solution of the
following initial value problem: 
\begin{equation*}
\left\{ 
\begin{array}{cc}
\left( Hu\right) \left( x,t\right) =0\text{,} & x\in K^{n}\text{, \ }t\in K,
\\ 
&  \\ 
u\left( x,0\right) =f_{0}\left( x\right) , & 
\end{array}%
\right.
\end{equation*}%
where $f_{0}\left( x\right) \in \mathbb{S}(K^{n})$. Then 
\begin{equation}
\left\Vert u\left( x,t\right) \right\Vert _{L^{\sigma }\left( K^{n+1}\right)
}\leq A\left\Vert f_{0}\left( x\right) \right\Vert _{_{L^{2}\left(
K^{n}\right) }}\text{,}  \label{est}
\end{equation}%
for $\frac{2\left( 1+\beta _{\phi }\right) }{\beta _{\phi }}<\sigma \leq
\infty $.
\end{theorem}

\begin{proof}
Since 
\begin{eqnarray*}
u\left( x,t\right) &=&\int\limits_{K^{n}}\Psi \left( t\phi \left( \xi
\right) +\left[ x,\xi \right] \right) \mathcal{F}f_{0}\left( \xi \right)
\left\vert d\xi \right\vert \\
&=&\int\limits_{Y}\Psi \left( \left[ \underline{x},\underline{\xi }\right]
\right) \mathcal{F}f_{0}\left( \underline{\xi }\right) d\mu _{Y,S}\underline{%
\left( \xi \right) }\text{,}
\end{eqnarray*}%
where \ \underline{$\xi $}$=\left( \xi ,\xi _{n+1}\right) \in K^{n+1}$, 
\underline{$x$}$=\left( x,t\right) \in K^{n+1}$, 
\begin{equation*}
Y=\left\{ \underline{\xi }\in K^{n+1}\mid \xi _{n+1}=\phi \left( \xi \right)
\right\} ,
\end{equation*}%
and $d\mu _{Y,S}=\Theta _{S}d\sigma _{Y}$, with $\Theta _{S}$\ the
characteristic function of a compact open set $S$ \ containing the support
of $\mathcal{F}f_{0}$. By applying Theorem \ref{th5}, replacing $n$ with $%
n+1,$ and dualizing, \ \ one gets 
\begin{equation}
\left\Vert u\left( x,t\right) \right\Vert _{L^{\sigma }\left( K^{n+1}\right)
}\leq A\left\Vert f_{0}\left( x\right) \right\Vert _{_{L^{2}\left(
K^{n}\right) }}\text{,}  \label{rf11}
\end{equation}%
where $\sigma =\frac{2\left( 1+\beta \right) }{\beta }$ is the dual exponent
of $\rho $\ in Theorem \ref{th5}, and $0\leq \beta <\beta _{\phi }$,
therefore (\ref{rf11}) is valid for $\frac{2\left( 1+\beta _{\phi }\right) }{%
\beta _{\phi }}<\sigma \leq \infty $.
\end{proof}

\subsection{Wave-type Equations with Homogeneous Symbols}

In the cases $\phi \left( \xi \right) =a_{1}\xi _{1}^{2}+\ldots +a_{n}\xi
_{n}^{2}$\ and $n=1$, $\phi \left( \xi \right) =\xi ^{d}$\ by using Remark %
\ref{Rem2} we have the following estimations for the solution of Cauchy
problem (1).

\begin{theorem}
\label{cor1} If $\phi \left( \xi \right) =a_{1}\xi _{1}^{2}+\ldots +a_{n}\xi
_{n}^{2}$, then 
\begin{equation*}
\left\Vert u\left( x,t\right) \right\Vert _{L^{\frac{2\left( 2+n\right) }{n}%
}\left( K^{n+1}\right) }\leq C\left\Vert f_{0}\left( x\right) \right\Vert
_{_{L^{2}\left( K^{n}\right) }}\text{.}
\end{equation*}
\end{theorem}

\begin{theorem}
\label{cor2}If $n=1$ and $\phi \left( \xi \right) =\xi ^{d}$, then 
\begin{equation*}
\left\Vert u\left( x,t\right) \right\Vert _{L^{2\left( d+1\right) }\left(
K^{2}\right) }\leq C\left\Vert f_{0}\left( x\right) \right\Vert
_{_{L^{2}\left( K\right) }}\text{.}
\end{equation*}%
In particular \ if $d=3,$ then 
\begin{equation*}
\left\Vert u\left( x,t\right) \right\Vert _{L^{8}\left( K^{2}\right) }\leq
C\left\Vert f_{0}\left( x\right) \right\Vert _{_{L^{2}\left( K\right) }}%
\text{.}
\end{equation*}
\end{theorem}

\end{document}